\documentclass[runningheads]{llncs}
\usepackage[T1]{fontenc}
\usepackage{amsmath}
\usepackage{amssymb}
\usepackage{tcolorbox}
\usepackage{colortbl}
\usepackage{marvosym}
\let\Letter\relax
\usepackage[misc]{ifsym}
\usepackage{multirow}
\usepackage{graphicx}
\usepackage{booktabs}
\usepackage{graphicx}    
\usepackage{subcaption}  
\usepackage{caption}     
\begin{document}
\title{An Efficient Interaction Human-AI Synergy System Bridging Visual Awareness and Large Language Model for Intensive Care Units}
\titlerunning{Human-AI Synergy System for ICU}
%
\author{
Yibowen Zhao\inst{1} \and
Yiming Cao\inst{2} \and 
Zhiqi Shen\inst{2,3}\textsuperscript{(\Letter)} \and
Juan Du\inst{4} \and
Yonghui Xu\inst{1} \and
Lizhen Cui\inst{1} \and
Cyril Leung\inst{2,3}}
\institute{School of software, Joint SDU-NTU Centre for Artificial Intelligence Research (C-FAIR), Shandong University, China \and
Alibaba-NTU Global e-Sustainability CorpLab (ANGEL), Nanyang Technological University, Singapore \and
Joint NTU-UBC Research Centre of Excellence in Active Living for the Elderly (LILY), Nanyang Technological University, Singapore \and
Department of Critical Care Medicine, Qilu Hospital of Shandong University, China 
}
\authorrunning{Y. Zhao et al.}
%
%
\maketitle             
\begin{abstract}
Intensive Care Units (ICUs) are critical environments characterized by high-stakes monitoring and complex data management. However, current practices often rely on manual data transcription and fragmented information systems, introducing potential risks to patient safety and operational efficiency. To address these issues, we propose a human-AI synergy system based on a cloud-edge-end architecture, which integrates visual-aware data extraction and semantic interaction mechanisms. Specifically, a visual-aware edge module non-invasively captures real-time physiological data from bedside monitors, reducing manual entry errors. To improve accessibility to fragmented data sources, a semantic interaction module, powered by a Large Language Model (LLM), enables physicians to perform efficient and intuitive voice-based queries over structured patient data. The hierarchical cloud-edge-end deployment ensures low-latency communication and scalable system performance. Our system reduces the cognitive burden on ICU nurses and physicians and demonstrates promising potential for broader applications in intelligent healthcare systems.

\end{abstract}
\section{Introduction}

\begin{figure*}[t]
\centering
\renewcommand{\dblfloatpagefraction}{.9}
\includegraphics[width=\textwidth]{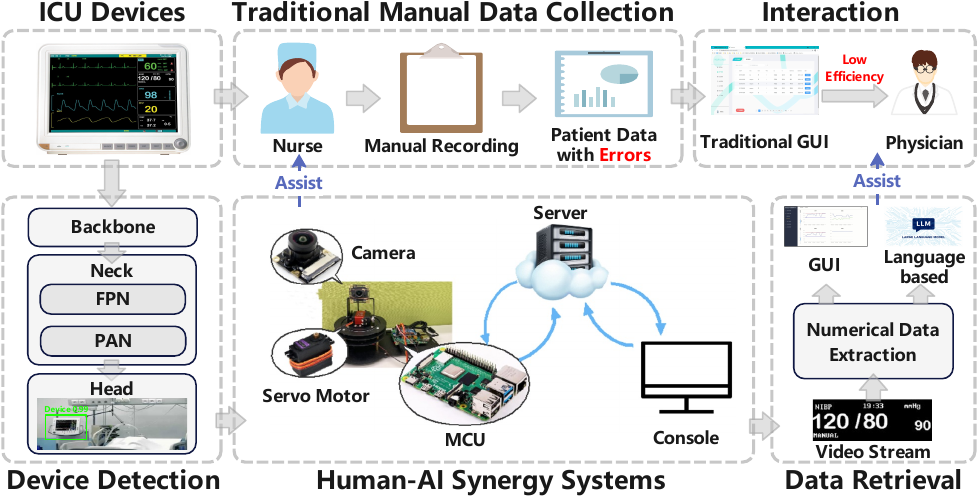}
\caption{Overview of the Proposed Human-AI Synergy System for ICU Monitoring. The system assists nurses by automatically capturing physiological data from bedside monitors via non-invasive screen recognition, and enhances physician interaction by enabling semantic-level query using a large language model interface.}
\label{fig:framework}
\end{figure*}

In the high-intensity and complex environment of the Intensive Care Unit (ICU), continuous patient monitoring and timely clinical decision-making are crucial but challenging~\cite{schlicht2024context}. Despite advances in digital health technologies, current clinical workflows are highly dependent on manual data entry and fragmented data retrieval that demand excessive manual intervention. Vital signs are typically dispersed across various standalone monitors, infusion pumps, ventilators, and dialysis machines. Each of these devices may display information independently on separate screens, necessitating frequent manual transcription into centralized systems~\cite{moy2021measurement,chen2022teamwork}.
From a system-level perspective, bedside devices in ICU lack standardized interoperability protocols, making automated data integration difficult without physical wiring or customized software drivers~\cite{ng2021network,gonzalez2021using}. 
Nurses in the ICU are required to manually document physiological measurements from multiple device screens every hour, creating substantial workloads prone to transcription errors~\cite{chen2022teamwork,moy2021measurement}. 
On the other side, physicians must manually integrate patient data across various isolated Graphical User Interfaces (GUIs) of Hospital Information System (HIS), which frequently leads to critical information omissions~\cite{hofmaenner2021use} and delayed clinical responses~\cite{moy2023understanding}.
Due to the reliance on physical connections, automating ICU data acquisition systems remains challenging in terms of deployment and updates, which significantly increases the cognitive load of medical personnel, error rates, and overall inefficiency.

In this paper, we focus on two core research questions that arise from these operational inefficiencies. \textbf{RQ1}: How can we design an automated visual information extraction system that captures and integrates real-time monitoring screen data from various ICU devices, without requiring physical device-level integration, to reduce the burden of nursing documentation and risk of errors~\cite{mosch2022creation,de2015information}? \textbf{RQ2}: How can we augment physicians’ interaction with the HIS interface through semantic-level natural language queries, so as to streamline the identification of critical clinical information and mitigate the risks associated with cognitive overload and omission?

To address the above research questions, we propose a collaborative cloud-edge-end human-AI synergy system, as illustrated in Figure~\ref{fig:framework}. The system leverages advanced visual computing techniques deployed on edge devices, eliminating the need for direct physical connections and facilitating seamless ICU deployment. 
The data from the ICU device screen is captured by video processing and the extracted patient physiological parameters are transmitted to the cloud system, reducing the workload of nurses.
The end devices of the system enhances physician interaction efficiency, streamlining data retrieval and decision support by integrating a Large Language Model (LLM)-powered GUI.
Specifically, our system incorporates two complementary AI-driven solutions.

To address transcription errors and enable real-time data aggregation, we develop an edge-deployed, visual-aware data extraction module for ICU screen recognition and physiological data capture. The edge devices integrate high-resolution cameras, a motorized gimbal, and a micro-computing unit (MCU) running a lightweight YOLOv5 model for detecting relevant screen regions. Once the display regions are identified, the edge system performs on-device Optical Character Recognition (OCR) in combination with Convolutional Recurrent Neural Networks (CRNN)~\cite{shi2016end}, extracting and digitizing numerical values directly from the monitor images. This pipeline eliminates the need to transmit raw video feeds by uploading only the structured and filtered patient data to the cloud, thereby reducing bandwidth consumption and enhancing privacy. To facilitate intuitive access to this structured data, we introduce a sophisticated human-AI interface powered by Large Language Models (LLMs). Leveraging prompt engineering techniques, the system dynamically interprets physicians’ natural language queries in conjunction with patient-specific information, enabling efficient and context-aware interaction with the underlying database. Rather than requiring manual scripting or pre-defined templates, the LLM module generates tailored responses and retrieves relevant clinical data through semantic understanding, without extensive computational overhead or retraining. By integrating visual perception with semantic understanding at the edge and cloud levels, our system provides a practical, scalable solution that enhances data integration, reduces clinician workload, and improves real-time interaction efficiency.

\section{Related Works}
In recent years, the critical care domain has seen increasing interest in leveraging intelligent technologies to address persistent challenges in ICU, such as data fragmentation, manual documentation burdens, and physician cognitive overload. In this section, we review previous works, with particular attention to two interrelated components: (1) device perception and data integration, and (2) semantic interaction and high-efficiency interface design.

\subsection{Device Perception and Data Integration in ICU Environments}
Efficient ICU monitoring requires reliable and timely access to multi-modal physiological data. However, most bedside devices lack unified standards, making integration non-trivial. Noteboom et al.~\cite{noteboom2025intensive} propose a structured pipeline to aggregate real-time data from monitors, enabling centralized analysis. Feng et al.~\cite{feng2024design} extend this paradigm with an AIoT-enabled ICU command center that synchronizes device signals for automated diagnostics and alerts. Although recent systems improve centralization, Gagliardi et al.~\cite{gagliardi2025assessing} report persistent incompatibility in hardware protocols and data formats across ICUs. Mosch et al.~\cite{mosch2022creation} address this issue by proposing an implementation framework centered on physician involvement and incremental digital integration. Building on this vision, Barroca Filho et al.~\cite{de2021iot} present an IoT-based system that aggregates wearable and fixed monitor data through wireless channels, proving practical under pandemic pressure. We adopt such minimal-disruption principle but shift toward a vision-based approach that bypasses device APIs entirely.

Visual capture offers an alternative pathway for real-time data extraction without device-level integration~\cite{rietveld2025let}. Nitayavardhana et al.~\cite{nitayavardhana2025streamlining} demonstrate a screen-reading system using OCR to transcribe vital signs with over 97\% accuracy, substantially reducing nursing workload. Jeon et al.~\cite{jeon2023romi} introduce ROMI, a mobile robot equipped with computer vision that autonomously reads ICU device screens. Froese et al.~\cite{froese2021computer} implement a continuous infusion pump capture module using OCR to integrate dosage data into monitoring dashboards. Coscolluela et al.~\cite{coscolluela2023using} present a low-cost mobile capture solution for use in under-resourced ICU settings. Together, these efforts validate visual recognition as a flexible, hardware-agnostic strategy.

Inspired by these findings, our system incorporates a non-invasive device perception module that combines real-time video capture with an optimized YOLOv5-based detector for identifying screen regions. Structured OCR is applied to extract vital signs, which are then normalized into standard medical concepts. This approach enables high-fidelity data integration across heterogeneous ICU environments without requiring any physical rewiring or vendor-specific access.

\subsection{Semantic Interaction and High-Efficiency Interface Design}
While acquiring structured data is critical, efficient interaction with that data is equally essential in high-pressure ICU settings. Recent work has shifted from predictive automation to enhancing physician-system interaction. Hyland et al.~\cite{hyland2020early} build models for circulatory failure prediction, yet Eini-Porat et al.~\cite{eini2022tell} argue that clinical impact also depends on the interpretability and accessibility of information. Gaber et al.~\cite{gaber2025evaluating} benchmark retrieval-augmented LLMs for ICU data summarization and triage, revealing their potential for improving clinical communication. Yang et al.~\cite{yang2025large} develop ICU-GPT, an LLM-based interface that auto-generates SQL queries and visualizations from natural-language inputs, enabling physicians to extract trends without scripting. Wu et al.~\cite{wu2025large} show that LLMs also support medical education by improving understanding and recall in diagnostic scenarios. These findings support our semantic interaction module, which adopts LLMs not for autonomous decision-making, but to enhance the speed, fluency, and semantic alignment of physician queries over patient data.

Beyond textual interfaces, voice interaction has emerged as a promising modality to further reduce cognitive and operational friction in ICU workflows. Peine et al.~\cite{peine2023standardized} introduce a speech-driven documentation assistant that significantly improves input speed and accuracy over manual entry. Building on this foundation, King et al.~\cite{king2023voice} develop a context-aware assistant that actively listens to clinical rounds and provides timely prompts aligned with ongoing discussions. Extending this line of research, we embed speech-based access into our interface layer, enabling physicians to retrieve patient states, compare measurements, and monitor temporal trends through intuitive, hands-free queries. To ensure system utility across diverse clinical settings, Hsu et al.~\cite{huo2023reducing} and Acharya et al.~\cite{bostan2024customizing} emphasize the need for adaptive interfaces that align with user profiles and varying workflow demands. Complementing this perspective, Poncette et al.~\cite{poncette2022remote} and Merola et al.~\cite{merola2025telemedicine} underscore the critical role of mobile-friendly, low-latency ICU dashboards—particularly in the context of staffing shortages or surge scenarios. These findings collectively highlight that interaction responsiveness and contextual appropriateness are key to physician engagement and system adoption.

By integrating visual data extraction with semantic understanding and speech-enabled interfaces, we provide a comprehensive solution that enhances information accessibility, minimizes interaction latency, and aligns naturally with ICU workflows. This unified architecture bridges non-invasive perceptual input with high-efficiency, multi-modal interaction, offering physicians a fluid and intuitive means of engaging with complex patient data in real time.

\section{Method}
In this section, we present the design of the proposed human-AI synergy system for ICU environments. Our approach integrates two core components: a \textbf{visual-aware data extraction module} deployed on edge devices to autonomously capture physiological data, and a \textbf{semantic interaction module} enhanced by a LLM for intuitive clinical data querying. The data extraction module consists of two submodules: (1) ICU device detection, which identifies relevant monitor regions using real-time visual analysis, and (2) data retrieval and structuring, which extracts numerical values via OCR and formats them into standardized clinical records. These components operate within a unified cloud-edge-end architecture, facilitating scalable deployment and seamless physician interaction.

\begin{figure}[t]
    \centering
    \includegraphics[width=0.85\linewidth]{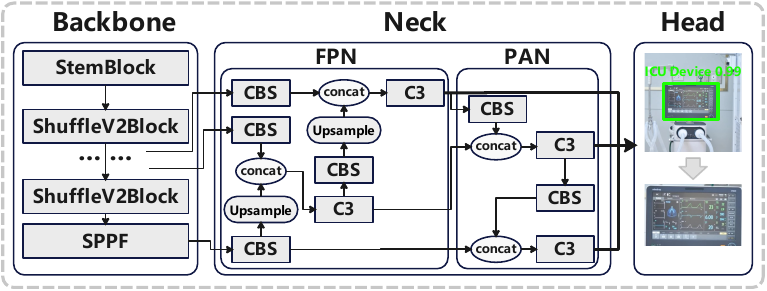}
    \caption{Architecture of Visual-Aware ICU Device Detection Module.}
    \label{fig:yolo}
\end{figure}

\subsection{ICU Device Detection}
To mitigate the challenges associated with manual transcription and intrusive hardware integration, we propose a non-invasive ICU device detection module designed to automatically identify regions on bedside monitors displaying patient physiological data. Instead of directly extracting numerical values at this stage, the module precisely detects and locates relevant screen areas, preparing data for subsequent processing.

The system integrates a servo-controlled, high-resolution camera with an edge-based detection pipeline, continuously scanning the ICU environment. Leveraging real-time object detection, the module accurately identifies screen regions without modifications to existing ICU hardware or proprietary interfaces, ensuring broad applicability and ease of deployment. Formally, given a video frame captured at time  \( t \), denoted as \( \mathbf{V}_t \in \mathbb{R}^{H \times W \times C} \), the detection objective is defined as:
\begin{equation}
\mathcal{B}_t = { b_j \in \mathbb{R}^4 \mid \text{score}(b_j) \geq \theta, j = 1, \dots, M }
\end{equation}
where each bounding box \( b_j \) represents the detected screen coordinates, and \( \theta \) is the confidence threshold.

We employ an optimized YOLOv5~\cite{huang2023classification,wang2022novel} model (illustrated in Figure~\ref{fig:yolo}), chosen for its robust performance in complex visual environments and suitability for embedded edge inference. The model comprises three primary components: the Backbone, Neck, and Head. The Backbone extracts hierarchical visual features using a Stem Block, multiple ShuffleNet version 2 (ShuffleV2) blocks for computational efficiency, and a Spatial Pyramid Pooling - Fast (SPPF) module for multi-scale feature aggregation. The Neck consists of a Feature Pyramid Network (FPN) and a Path Aggregation Network (PAN), designed to enhance semantic representation across different scales. It incorporates several key operators such as Convolution-BatchNorm-SiLU (CBS) blocks, Cross Stage Partial (C3) modules for efficient feature reuse, and Upsample layers to refine spatial resolution during top-down feature fusion. Finally, the Head outputs dense bounding box predictions with associated confidence scores, followed by Non-Maximum Suppression (NMS) to filter redundant detections and retain the most probable screen regions.

This localized detection module forms the front-end of the data acquisition system. Its modular, hardware-agnostic design enables rapid deployment in real-world ICU while ensuring reliable and consistent screen localization performance under diverse environmental conditions.

\begin{figure}[t]
    \centering
    \includegraphics[width=0.85\linewidth]{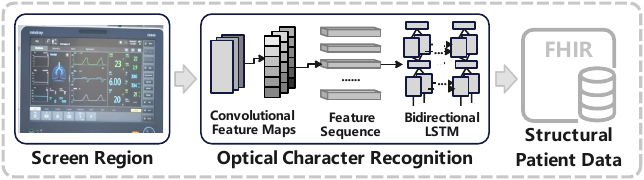}
    \caption{OCR Pipeline for ICU Screen Text Digitization and FHIR Structuring.}
    \label{fig:ocr}
\end{figure}

\subsection{Data Retrieval and Structuring}
Subsequent to identifying screen regions \( \mathcal{B}_t \), the data retrieval module extracts, interprets, and structures numerical physiological data from these screen captures. Departing from traditional hardware-integrated methods, our system employs a contactless digitization pipeline based on OCR and layout-aware segmentation heuristics, ensuring robust and adaptable data capture.

Each detected region \( \mathbf{I}_{b_j} \) is analyzed using vertical-horizontal projection profiling and connected-component analysis to isolate semantically coherent key-value pairs (e.g., 'HR: 85', 'SpO\textsubscript{2}: 98'). These candidate text clusters are then processed by an OCR pipeline, implemented using a Convolutional Recurrent Neural Network (CRNN)~\cite{shi2016end}. The CRNN architecture consists of three stages, as presented in Figure~\ref{fig:ocr}: (1) Convolutional layers generate low-level feature maps from the input screen region; (2) these maps are converted into a sequential feature representation through column-wise flattening; and (3) the sequence is passed through a Bidirectional Long Short-Term Memory (BiLSTM) network to capture temporal dependencies in both forward and backward directions. Finally, a Connectionist Temporal Classification (CTC) decoder generates recognized text strings without explicit character-level alignment:
\begin{equation}
\hat{v}_j = f_{\text{ctc}}(\text{BiLSTM}(\text{CNN}(\mathbf{I}_{b_j}^{(i)})))
\end{equation}
Results are then normalized to standard clinical concepts and measurement units:
\begin{equation}
(k_j, u_j) = f_{\text{map}}(\hat{v}_j)
\end{equation}
The final structured data output at time $t$, encoded in Fast Healthcare Interoperability Resources (FHIR) format, is:
\begin{equation}
\mathcal{D}_t = { (k_j, \hat{v}_j, u_j) \mid j = 1, \dots, M }
\end{equation}
This structured data retrieval module transforms raw visual data into standardized clinical records, enabling effective semantic-level interaction and analysis. The resulting structured data is then securely transmitted to the cloud, where it can be further integrated and queried by downstream semantic modules.

\subsection{Semantic Interaction via LLM-Augmented Interface}

Effective clinical decision-making requires rapid, intuitive access to patient data, often impeded by traditional fragmented ICU interfaces. Our semantic interaction module addresses this challenge by integrating automatic speech recognition (ASR) with a LLM at the cloud endpoint, enabling physicians to perform natural language queries. Formally, given structured patient records $\mathcal{D}_{1:T}$ and a spoken query transcribed to text $q$, the LLM produces clinically relevant responses:
\begin{equation}
a = \text{LLM}(q; \mathcal{D}_{1:T}, \mathcal{P})
\end{equation}
where $\mathcal{P}$ includes task-specific prompt templates embedding clinical context and temporal reasoning instructions. The prompt example is presented as follows:
\begin{tcolorbox}[colback=gray!5!white,colframe=black!30,title=Prompt Example: Temporal Trend Query with Patient Context]
\small
\textbf{Patient Information:}\\
Age: 72, Gender: Male, Diagnosis: COPD (Chronic Obstructive Pulmonary Disease), Past Medical History: Hypertension, Ex-smoker\\[0.5ex]

\textbf{Input:} ICU vital signs over the past 6 hours:\\
\texttt{[T0: HR 105 bpm, RR 24 bpm, SpO\textsubscript{2} 96\%], ..., [T6: HR 118 bpm, RR 28 bpm, SpO\textsubscript{2} 93\%]}\\[0.5ex]

\textbf{Query:} Has the patient's respiratory rate become increasingly unstable in the last 2 hours?\\[0.5ex]

\textbf{Instruction:} Based on the patient's known COPD history and provided vital signs, analyze whether there is a trend of increasing instability in respiratory rate. Identify any abnormal fluctuations, sudden spikes, or deviation from expected ranges. Report your findings with relevant timestamps. Keep the response concise and clinically interpretable.
\end{tcolorbox}

This approach provides concise, evidence-based answers and visualizes critical data points underlying clinical insights, thus significantly reducing cognitive load and facilitating efficient decision-making. By integrating semantic interaction capabilities, this module substantially enhances physicians’ ability to access patient data efficiently, enabling informed and timely clinical decisions.

\begin{figure}[t]
    \centering
    \includegraphics[width=0.6\linewidth]{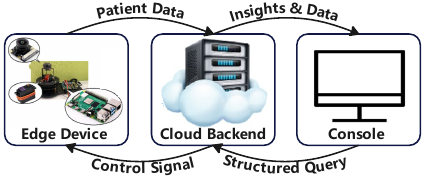}
    \caption{System-Level Architecture of Cloud-Edge-End Human-AI Synergy Platform}
    \label{fig:sys}
\end{figure}

\subsection{System Integration}
To enable practical deployment in ICU environments, the proposed system adopts a cloud-edge-end hybrid architecture, as presented in Figure~\ref{fig:sys}, that balances local processing efficiency with centralized semantic intelligence. The edge layer is composed of embedded visual agents—camera-mounted edge units deployed at the patient bedside—that continuously capture device screens and locally perform screen region detection and OCR-based value extraction. These operations are optimized for low-latency execution and designed to function without physical integration with medical devices, thereby preserving compatibility across heterogeneous clinical infrastructure.

Extracted physiological data streams are securely transmitted to a cloud-based coordination center via encrypted channels. The cloud layer hosts the structured data warehouse and serves as the execution environment for the LLM-based semantic engine. This separation of concerns enables scalable inference on multi-patient records, multi-modal query handling, and federated model updates without burdening the edge devices. To support seamless physician interaction, the system integrates a lightweight, voice-activated assistant into the GUI, available on both hospital terminals and mobile devices. This assistant supports wake-word activation, speech-to-text transcription, and audio playback of LLM-generated responses, enabling intuitive, hands-free interaction. All queries and responses are logged and timestamped to ensure transparency, traceability, and clinical safety compliance. This integration pipeline keeps the edge modules lightweight and privacy-preserving, while the cloud services handle computationally intensive tasks such as semantic reasoning, temporal trend abstraction, and natural language generation. In the event of network failure, edge modules are designed to fallback to local buffering and partial inference, ensuring robustness and continuity in deployment.

Overall, this architecture supports real-time, multimodal monitoring and voice-based clinical interaction in a modular and deployable manner, closing the loop between automated sensing, semantic understanding, and human decision-making in intensive care environments. 

\section{System Implementation and Interaction Examples}
In this section, we present the system implementation for ICU monitoring scenarios from two key perspectives: edge device integration and intelligent interaction interface design. The following subsections detail the hardware architecture of the edge devices and the design of the user interface.

\subsection{Simulation and Video-Capture Device Architecture}
\begin{figure*}[t]
    \centering
    \begin{subfigure}[b]{0.5\textwidth}
        \centering
        \includegraphics[width=\textwidth]{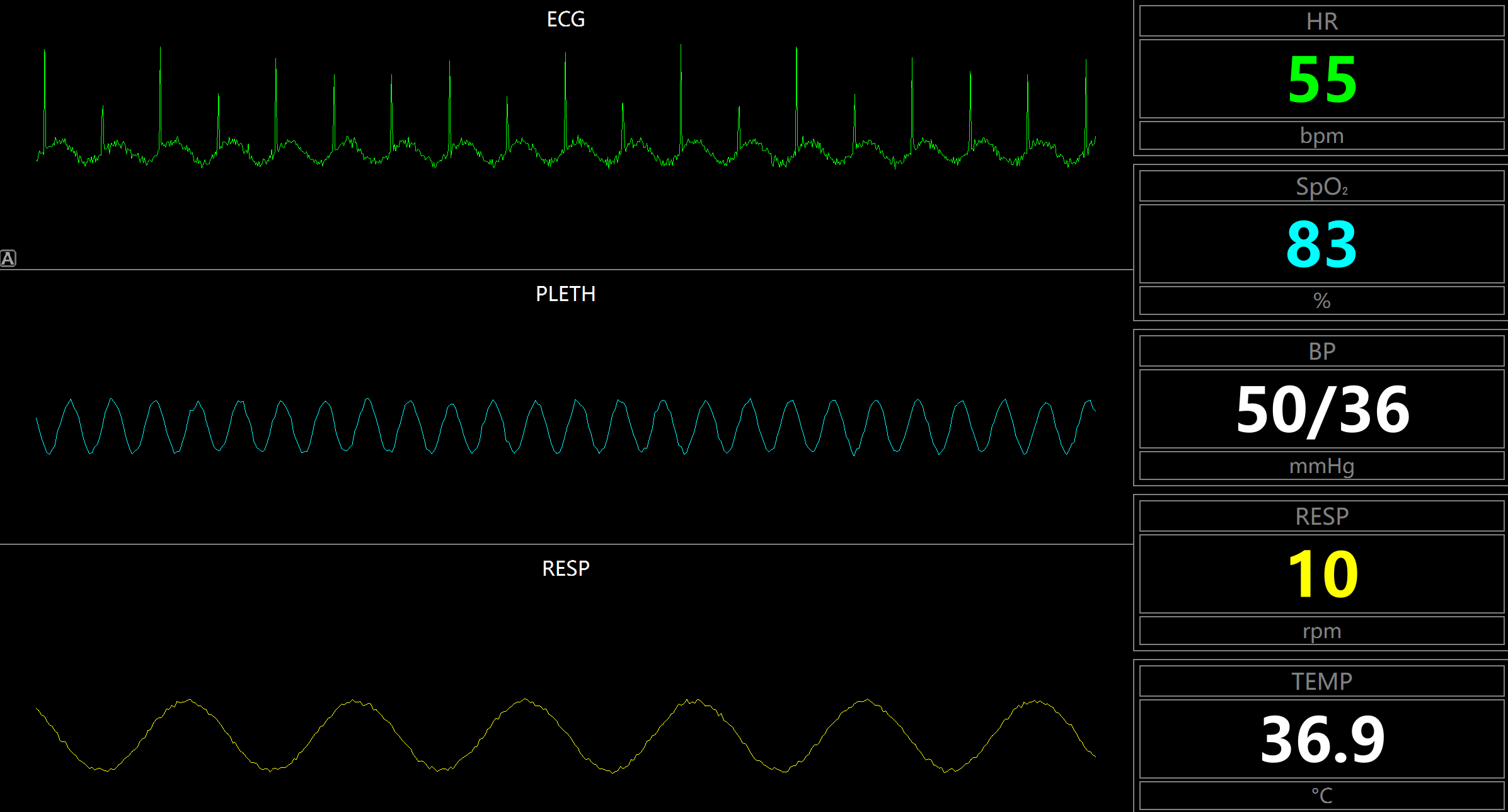}
        \caption{Simulated ICU device screen}
        \label{fig:simulated_monitor}
    \end{subfigure}
    \hfill

    \begin{subfigure}[b]{0.49\textwidth}
        \centering
        \includegraphics[width=\textwidth]{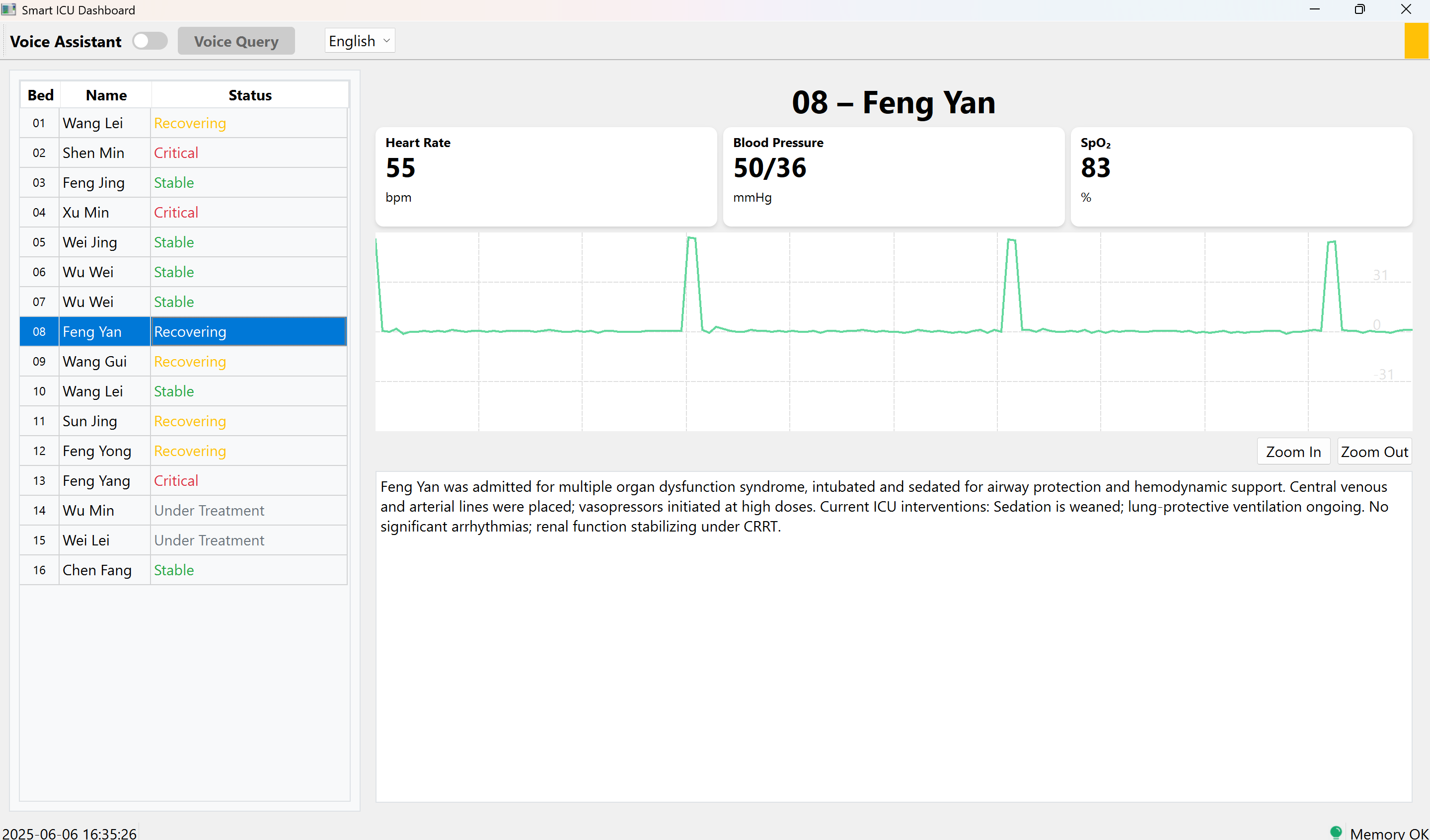}
        \caption{GUI (English)}
        \label{fig:gui_en}
    \end{subfigure}
    \hfill
    \begin{subfigure}[b]{0.49\textwidth}
        \centering
        \includegraphics[width=\textwidth]{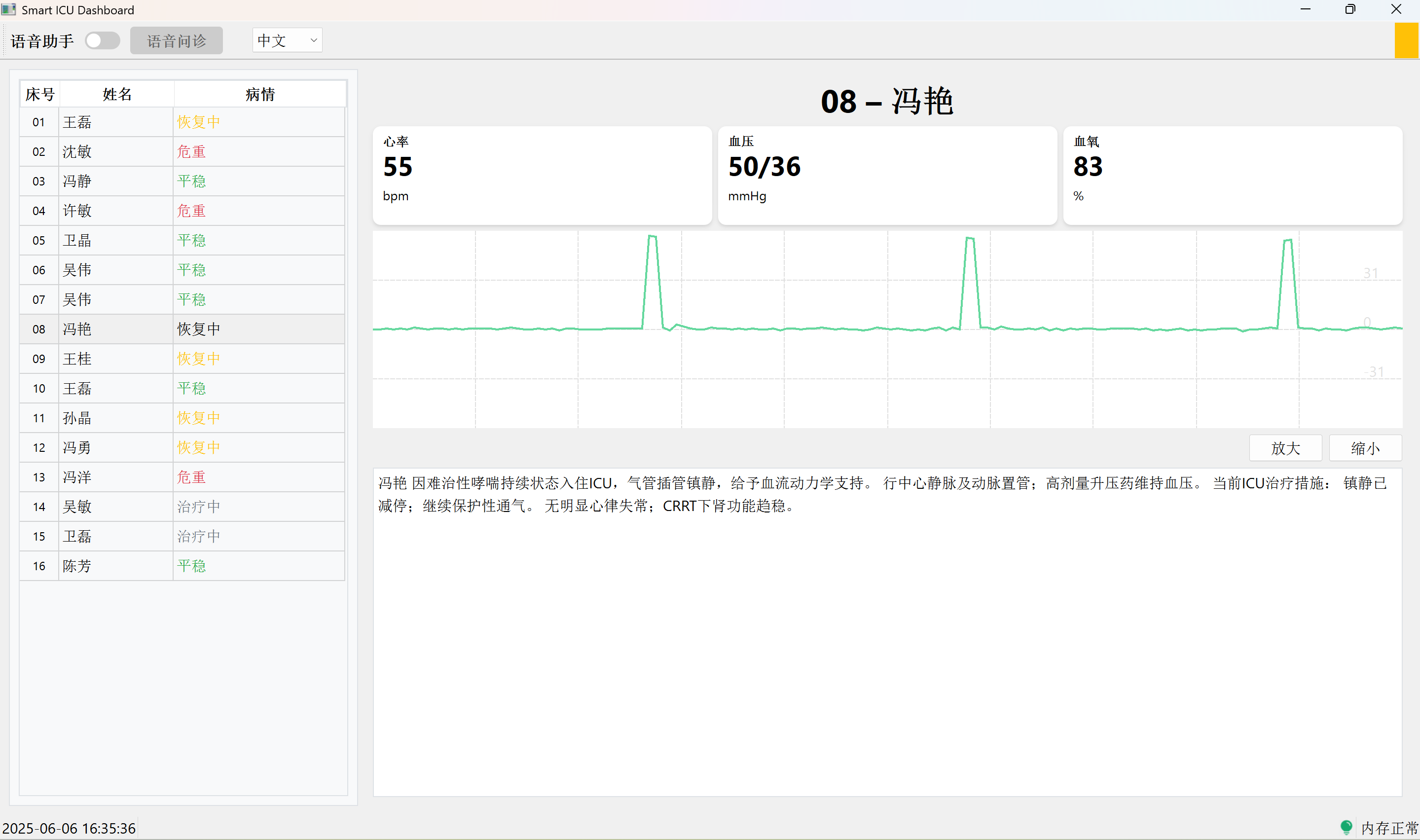}
        \caption{GUI (Chinese)}
        \label{fig:gui_zh}
    \end{subfigure}
    \caption{Integrated simulation and bilingual GUI environment for ICU monitoring, combining device interface capture with English and Chinese clinical dashboards.}
    \label{fig:combined_interface}
\end{figure*}

To facilitate realistic demonstrations, we design a simulated ICU bedside physiological monitoring setup (as illustrated in Figure~\ref{fig:simulated_monitor}) integrated with an edge-based video-capture device. This device includes a Raspberry Pi 4 Model B microcontroller unit (MCU), an OV9726 camera module, and MG90S servo motors precisely controlled by a PCA9685 pulse-width modulation (PWM) driver. Powered by a quad-core Cortex-A72 processor, the Raspberry Pi performs real-time YOLOv5-based detection and OCR, effectively minimizing system latency and reducing bandwidth by performing key inference operations locally. The servo-driven camera mechanism enables dynamic orientation adjustments, compensating for variable monitor placements and potential occlusions. The OV9726 camera, selected for its compact form factor and low power consumption, provides high-resolution (1280×800 pixels) color video at 60 frames per second, with a wide 70° horizontal field of view. It reliably captures ICU monitor outputs from a working distance of 30-50 cm. Together, these components enable robust real-time video processing and accurate extraction of numerical and waveform data critical for ICU monitoring tasks.

\subsection{Graphical User Interface Design}
The GUI serves as the central platform for physician interaction, designed to present real-time patient data in a clear and organized manner, with support for multilingual operation—English (Figure~\ref{fig:gui_en}) and Chinese (Figure~\ref{fig:gui_zh}). The interface comprises two primary panels: a patient overview panel and a detailed information panel. The overview panel continuously updates and categorizes patient statuses (e.g., Critical, Under Treatment, Recovering, Stable), clearly displaying bed identifiers and key demographic details. Upon selecting a patient, the detailed panel is dynamically updated with real-time physiological indicators, including graphical gauges for vital signs such as heart rate, blood pressure, and oxygen saturation (SpO$_2$), refreshed every second to support timely clinical assessment. Below these gauges, an interactive electrocardiogram (ECG) waveform component provides continuous signal visualization, supporting zoom and scroll functions for detailed inspection. Patient demographic and medical history data are shown in a concise, read-only format. Integrated into the GUI is a voice-assistant module, accessible via a toggle button in the top toolbar. This feature enables physicians to perform hands-free, natural language queries, processed in real time by a LLM that draws on current patient data. The system supports multilingual responses, which can either be audibly delivered in the selected language or displayed in concise textual form directly within the GUI.

\begin{table}[t]
\centering
\caption{Representative Queries and LLM‐Generated Answers in an ICU Scenario}
\label{tab:llm_qa_examples}
\begin{tabular}{p{4.0cm} | p{7.9cm}}
\toprule
\multicolumn{1}{c|}{\textbf{Query}} & \multicolumn{1}{c}{\textbf{Answer}} \\
\midrule
What is the current heart rate of the patient in Bed 03? & The current heart rate of the patient in Bed 03 is 106 bpm, measured at 14:22. \\

\hline
Has the patient’s SpO\textsubscript{2} dropped below 90\% in the past hour? & Yes. The SpO\textsubscript{2} dropped to 88\% at 13:17 and remained below 90\% for approximately 5 minutes before rising to 91\% by 13:23. \\

\hline
Compare the respiratory rate now and two hours ago. Has it increased? & The current respiratory rate is 26 bpm, up from 20 bpm at 12:22, indicating a moderate increase over the two‐hour interval. \\

\hline
Is there any abnormal fluctuation in blood pressure within the last 30 minutes? & Yes. The systolic blood pressure spiked from 118 mmHg at 13:45 to 152 mmHg at 14:00, followed by a drop to 110 mmHg at 14:10. Such variation may suggest hemodynamic instability. \\

\hline
Summarize the trend of the patient’s vital signs over the last 6 hours. & Over the past 6 hours, heart rate rose gradually from 92 bpm at 12:00 to 112 bpm at 18:00, respiratory rate increased from 20 to 27 bpm, and SpO\textsubscript{2} fell from 97\% to 94\%. These patterns may indicate early respiratory decompensation. \\

\hline
What is the average heart rate over the past 2 hours, and does it exceed 100 bpm? & The average heart rate between 16:00 and 18:00 is 102 bpm, which exceeds 100 bpm and meets the clinical threshold for tachycardia in this COPD patient. \\
\bottomrule
\end{tabular}
\end{table}

\subsection{Representative Query-Answer Examples}
To demonstrate the clinical utility of the proposed system, we present representative query-response interactions between physicians and the LLM-powered semantic interface in Table~\ref{tab:llm_qa_examples}. These examples illustrate the LLM’s ability to synthesize complex physiological signals and patient-specific trends into clear, clinically actionable insights. The structured responses concisely summarize vital sign abnormalities, temporal fluctuations, and deviations from expected clinical thresholds, thereby reducing physician cognitive load and facilitating timely, informed decision-making. These examples substantiate the effectiveness of our system in enhancing clinical efficiency, decision accuracy, and semantic accessibility in high-pressure ICU environments.

\section{Conclusion}
This paper proposes a novel human-AI synergy system explicitly designed for the demanding operational context of ICU. Our integrated framework mitigates manual data entry burdens through an advanced visual detection and OCR-based extraction pipeline, deployed on edge devices to ensure broad compatibility and low latency. Furthermore, the incorporation of a LLM significantly enhances physician interaction by enabling intuitive, semantic-level queries via speech. The modular cloud-edge-end architecture facilitates robust and scalable deployment across diverse clinical environments. Clinical trials with real ICU deployments are planned in future work to validate usability and effectiveness. Future research will also focus on enhancing system adaptability for broader medical applications, aiming to promote widespread adoption and deliver substantial clinical impact.

\section*{Acknowledgement}
This research is supported, in part, by Alibaba Group and NTU Singapore through Alibaba-NTU Global e-Sustainability CorpLab (ANGEL).

\bibliographystyle{splncs04}
\bibliography{hcii}

\end{document}